\documentstyle[12pt,epsf]{article}
\newcommand\pubnumber{SLAC-PUB-8021}
\newcommand\pubdate{September, 1999}

\def\Title#1{\begin{center} {\Large #1 } \end{center}}
\def\Author#1{\begin{center}{ \sc #1} \end{center}}
\def\Address#1{\begin{center}{ \it #1} \end{center}}

\def\submit#1{\begin{center}Submitted to {\sl #1} \end{center}}
\def\doeack{\footnote{Work supported by the Department of Energy,
                     contract DE--AC03--76SF00515.}}
\def\SLAC{Stanford Linear Accelerator Center\\
    Stanford University, Stanford, California 94309 USA}
\newcommand\pubblock{\rightline{\begin{tabular}{l} \pubnumber\\
         \pubdate  \end{tabular}}}
\newenvironment{Abstract}{\begin{quotation} \begin{center}
                       ABSTRACT
     \end{center}\bigskip  }{\end{quotation}}
\def\beq{\begin{equation}}
\def\eeq#1{\label{#1}\end{equation}}
\def\eeqn{\end{equation}}
\def\beqa{\begin{eqnarray}}
\def\eeqa#1{\label{#1}\end{eqnarray}}
\def\eeqan{\end{eqnarray}}
\def\CR{\nonumber \\ }
\def\leqn#1{(\ref{#1})}
\def\Acknowledgements{\bigskip  \bigskip \begin{center} \begin{large}
             \bf ACKNOWLEDGEMENTS \end{large}\end{center}}
\let\bar=\overbar
\def\Dslash{\not{\hbox{\kern-4pt $D$}}}
\def\dslash{\not{\hbox{\kern-2pt $\del$}}}
\def\half{\frac{1}{2}}
\def\thalf{\frac{3}{2}}

\def\L{{\cal L}}

\def\del{\partial}

\def\mw{m_W}
\def\mt{m_t}

\def\msb{{\bar{\ssstyle M \kern -1pt S}}}

\def\ch#1{\widetilde\chi^+_{#1}}
\def\chm#1{\widetilde\chi^-_{#1}}

\def\s#1{\widetilde{#1}}

\def\etal{{\it et al.}}

\begin{document}
\begin{titlepage}
\pubblock

\vfill
\Title{Scalar Top Quark as the Next-to-Lightest Supersymmetric Particle}
\vfill
\Author{Chih-Lung Chou\footnote{Present address:
      Institute of Physics, Academia Sinica, Nankang, Taipei 11529, ROC}
            and Michael E. Peskin\doeack}
\Address{\SLAC}
\vfill
\begin{Abstract}
We study phenomenologically the scenario in which the scalar top quark
is lighter than any other standard supersymmetric partner and also 
lighter than the top quark, so that it decays to the gravitino via $\s t \to
W^+ b \s G$. In this case, scalar top quark events would seem
to be very difficult to separate from top quark pair production.
However, we show that, even at a hadron collider, it is possible to 
distinguish these two reactions.  We show also that the longitudinal 
polarization of the final $W^+$ gives insight into the scalar top and
wino/Higgsino mixing parameters.
\end{Abstract}
\medskip
\submit{Physical Review {\bf D}}

\vfill
\end{titlepage}
\def\thefootnote{\fnsymbol{footnote}}
\setcounter{footnote}{0}
\section{Introduction}

Supersymmetry has been studied for a long time as the possible framework for
elementary particle theories beyond the standard 
model \cite{Nilles, MSSM, Dawson}.
  It provides a natural solution
to the hierarchy problem,  allowing a small value, in fundamental terms, 
for the weak interaction scale.   It also 
allows the measured values of the standard model coupling constants to be
consistent with grand unification.  Still, if  Nature is supersymmetric, 
some new interaction must spontaneously break supersymmetry and transmit
this information to the supersymmetric partners of the standard model 
particles.   Two different approaches have been followed to model
supersymmetry breaking.  The first is the idea that supersymmetry
breaking is transmitted by gravity and supergravity interactions \cite{SUGRA}.
In these scenarios, the 
supersymmetry breaking scale $\sqrt{F}$ is of  the order of $10^{11}$ GeV.
This large value implies that gravitino interactions are extremely weak,
and that the gravitino has a mass of the same size as the other supersymmetric
partners.  In this class of models,
 the lightest supersymmetric particle (LSP),
which is the endpoint of all superpartner decays, is most often taken to 
be the superpartner of the photon, or, more generally, a neutralino.

The second  approach 
 uses the gauge interactions to transmit the information of supersymmetry
breaking to the standard model 
partners \cite{GMSB,GMSBtwo,GMSBthree}.  In these 
gauge-mediated scenarios, the supersymmetry-breaking scale $\sqrt{F}$ is
typically much smaller than in the gravity-mediated case, so that the 
gravitino $\s G$ 
is almost always the LSP.  All other superpartners are unstable
with respect to decay to the gravitino, though sometimes with a
lifetime long on the time scale relevant to  collider physics.

In gauge-mediated scenarios, direct decay to the gravitino is hindered 
by a factor $1/F$ in the rate.  Thus, attention shifts to those particles
which have no allowed decays except through this hindered mode.  Such a 
particle is called a next-to-lightest supersymmetric particle (NLSP).
Any of the typically light superpartners can play the role of the NLSP, and
the collider phenomenology of a given model depends on which 
is chosen.
For example,
 if the gaugino-like lightest neutralino $\tilde \chi^0$ is the NLSP and 
decays inside the collider, supersymmetry reactions will end with  the 
decay
 ${\s \chi^0}\rightarrow \gamma {\s G}$, producing a direct photon
plus missing energy.  Other common choices for the
NLSP are the lepton partners and the Higgs boson.
More involved scenarios are also possible \cite{ltotau}.

In this paper, we consider the possibility that the 
 the lightest scalar top quark (stop, or  $\s t_1$)
is the NLSP of a  gauge-mediation scenario \cite{help}.
  It is typical in supersymmetric
models that the stop receives negative radiative corrections to its
mass through its coupling to the Higgs sector.  In addition, 
the mixing between the partners of the $t_L$ and $t_R$ is typically
sizable, and this drives down
 the the lower mass eigenvalue. It is not uncommon in models that the 
lighter stop is lighter than the top quark, and it is possible to 
arrange that it is also lighter than the sleptons and 
charginos \cite{stopone, stoptwo}.
The existence of this possibility, though, poses a troubling question
for experimenters.  In this scenario, the dominant decay of the lighter 
stop would be the three-body decay $\s t \to b W^+ \s G$.  The $\s G$ is 
not observable, and the rest of the reaction is extremely similar to the
standard top decay $t \to W^+ b$.  The cross section for stop pair
production is smaller than that for top pair production at the same mass.
Thus, it is possible that the top quark events discovered at the Tevatron
collider contain stop events as well.  How could we ever know?
In this paper, we address that question.

Our strategy will be to systematically analyze the three-body stop decay.
This decay process is rather complex, since the $\s G$ can be radiated
from the partners of $t$, $b$, or $W$, and since both the top and the $W$
partners can be a mixture of weak eigenstates.  For the application to the
Tevatron, one must take into account that the center-of-mass energy of the
production is unknown, and that the detectors can measure only a subset of
the possible observables.  Nevertheless, we will show that two observables
available at the Tevatron can cleanly distinguish between top and stop
events.  The first of these is the mass distribution of the observed 
$b$ jet plus lepton system which results from a leptonic $W$ decay.  The
second is the $W$ longitudinal polarization.  We will show that the first
of these observables gives a reasonably model-independent signature of 
stop production, while the second is wildly model-dependent and can be used
to gain insight into the underlying supersymmetry parameters.

This paper is organized as follows:  In Section 2, we set up our basic
formalism and state our assumptions.  In Section 3, we analyze the stop
decay rate and the $bW$ and $b\ell$ mass distributions.  In Section 4,
we present the $W$ longitudinal polarization in various models.
Section 5
gives our conclusions.

\section{Formalism and assumptions}

In this section, we define our notation and set out the  assumptions 
we will use in analyzing the stop decay process.   Our calculation will
be done within the framework of the minimal supersymmetric standard model
(MSSM) with R-parity conservation.  We will not consider any exotic 
particle other than those required in the MSSM.

 Our central assumption
will be that the lighter stop mass eigenstate $\s t_1$ is lighter than the 
top quark and also lighter than the  charginos and the $b$ superpartners,
while the gravitino is very light, as in gauge-mediation scenarios.
Under these assumptions, what would otherwise be the dominant decay
$\s t_1 \to t \s G$ is forbidden kinematically, so that the dominant
stop decay must proceed either by $\s t_1 \to b W^+ \s G$ or by
$\s t_1 \to c \s G$.  In the MSSM without additional flavor violation,
quark mixing angles suppress the decay to $c$ by a factor $10^{-6}$.  That
suppression makes this decay unimportant except near the boundary of 
phase space where $m \approx \mw$.  For this reason, we will ignore that 
decay in the rest of the paper.

 If the mass of the $\s t_1$ were larger than the mass of the
top quark, the $\s t_1$ would decay entirely through $\s t_1 \to t \s G$.
All observable characteristics of this decay would be exactly those of
top quark pair production, except that the two emitted gravitinos would lead
to a small additional transverse boost. 
 For such a heavy stop, the 
production cross section is less than 10\% of that for top quark pair
production.   Nevertheless, this process might be recognized from the 
fact that the top quark and antiquark would be given a small 
preferential polarization,
for example, in the $t_R \bar t_L$ helicity states if the $\s t_1$ is 
dominantly the partner of $t_R$.  The methodology of the  top polarization
measurement has been discussed in detail in the literature \cite{toppol},
so we will not analyze this case further here.

To analyze the case in which $\s t_1$ is lighter than the top quark, 
we begin by considering
the form of the scalar top quark mass matrix.  Including
the effects of soft breaking masses, Yukawa couplings, trilinear 
scalar couplings, and D terms, this matrix can be written in the 
$\s t_R$, $\s t_L$ basis as
\beq
M^2_{\tilde t} = \pmatrix{
   	m^2_{\tilde t_{R}} &  m_t(A_t+\mu \cot{\beta}) \cr
	m_t(A_t+\mu \cot{\beta})  & m^2_{\tilde t_{L}}  \cr} \ ,
\eeq{tmassmatrix}
where $A_t$, $\mu$, $m_t$, and $tan \beta$ denote, respectively,
 the trilinear 
coupling of Higgs scalars and sfermions, the supersymmetric Higgs mass term,
 the top quark mass, and the ratio of the two Higgs vacuum expectation 
values. The masses $m^2_{\tilde t_R}$ and  $m^2_{\tilde t_L}$ arise 
from the soft breaking, the D term contribution, and the top Yukawa 
coupling as follows:
\beqa
m^2_{\tilde t_R}&=& m^2_{\tilde U_3}+m^2_t + {2\over 3}\sin^2
{\theta_w} m^2_Z \cos 2{\beta} 
 \CR
m^2_{\tilde t_L}&=& m^2_{\tilde Q_3}+m^2_t + ({1 \over 2}-{2 \over 3} 
\sin^2{\theta_w}) m_Z^2 \cos2{\beta} \ , 
\eeqa{tmasses}
 where $\theta_w$ denotes the weak mixing angle and $m_Z$ is the $Z^0$ 
boson mass. The soft breaking masses $m^2_{\tilde U_3}$ and 
$m^2_{\tilde Q_3}$ are more model-dependent.   In many models, these 
masses are derived from flavor-blind  mass contributions by adding 
the effects of radiative corrections due to the top-Higgs Yukawa coupling
$\lambda_t$.
These corrections have the form
\beq
m^2_{\tilde U_3} \sim m^2_{\tilde U} - 2 {\lambda_t^2} {\tilde I}  \ , \qquad
m^2_{\tilde Q_3} \sim m^2_{\tilde Q} - {\lambda_t^2} {\tilde I},
\eeq{QRGE}
where the function $\tilde I$ denotes a  one-loop integral.
 The extra factor 2 in the expression for 
 the $m^2_{\tilde U_3}$ is due to 
the fact that loop diagram contains the $Q$ and Higgs isodoublets.
From this effect, we expect that $m^2_{\tilde U_3} < m^2_{\tilde Q_3}$.
One should note that there is a flavor-universal positive mass correction
due to diagrams with a gluino which combats the negative correction in 
\leqn{QRGE}.

The lightest stop mass eigenstate $\s t_1$ and its mass ${\s m}^2$ 
are  easily obtained by diagonalizing
the stop mass matrix \leqn{tmassmatrix}.  One finds
\beqa
 {\tilde t_1} &=& \cos{\theta_t} {\tilde t_L}
                  + \sin{\theta_t} {\tilde t_R}\CR
 {\tilde t_2} &=& \ sin{\theta_t} {\tilde t_L}
               -\cos{\theta_t} {\tilde t_R}\CR
{\s m}^2 &=& {1\over 2}\{ m^2_{\tilde t_R}+m^2_{\tilde t_L}
         - \sqrt{(m^2_{\tilde t_L}-m^2_{\tilde t_R})^2
                   +4m_t^2(A_t+\mu \cot\beta)^2}\}\CR
 tan{\theta_t}&=& -{{m_t (A_t+\mu \cot\beta)}
              \over {(m^2_{\tilde t_R} - m^2_1)}} \ .
\eeqa{tonemass}
In these formulae, $\theta_{\tilde t}$ denotes the stop
 mixing angle and is chosen to be in the range $-\pi/2 \le \theta_{\tilde t}
 \le \pi/2$.  The relations \leqn{tonemass} demonstrate the two 
mechanims mentioned in the introduction for obtaining a small value of 
  $m_{\tilde t_1}$:  First, the radiative correction \leqn{QRGE} could
be large due to the large value of $\lambda_t$; second, the left-right 
mixing could be large due a large value of $A_t$.  From here on, however,
we will take $\s m$ and $\theta_t$ to be phenomeonological parameters
to be determined by experiment.

Since the final state of the three-body $\s t_1$ decay includes the $W^+$, 
our analysis must include the supersymmetric partners of $W^+$ and $H^+$, 
the charginos.  In the MSSM, these states are mixtures of the winos
$\s{w}^\pm$ and the Higgsinos $h^\pm$.  In two-component fermion notation,
the left-handed chargino fields are written
\beq 
         \s{C}^+_i = (\s{w}^+ , ih^+)\ , \qquad 
  \s{C}^-_i = (\s{w}^- , ih^-) \ .
\eeq{charginos}
In this basis, the chargino mass matrix is 
\beq
         M_+  = \pmatrix{ m_2 &   -\sqrt{2}\mw \sin\beta \cr
                 -\sqrt{2}\mw \cos\beta  & \mu  } \ ,
\eeq{cmatrix}
where $m_2$ is the soft breaking mass of the $SU(2)$ gaugino, and $\mu$ is
the supersymmetric Higgs mass.  The matrix $M_+$ is diagonalized by 
writing $M_+ = (V_-)^T D V_+$, where $V_+$, $V_-$ are unitary; then the 
mass eigenstates are given by 
\beq
       \ch{i} = V_{+ij} \s{C}^+_j \ , \qquad  \chm{i} = V_{-ij} \s{C}^-_j \ .
\eeq{chdiag}
To be consistent with the assumption that the $\s{t}_1$ is the NLSP, we will 
consider only sets of parameters for which the mass of the  $\s{t}_1$ is
lower than either of the eigenvalues of $M_+$.

We analyze the couplings of superparticles to the gravitino by using the 
supersymmetry analogue of Goldstone boson equivalence.  The gravitino obtains
mass through the Higgs mechanism, by combining with the Goldstone 
fermion (Goldstino) associated with spontaneous supersymmetry breaking.
When the gravitino is emitted with an energy high compared to its mass, 
the helicity $h= \pm\thalf$ states come dominantly from the gravity multiplet
and are produced with gravitational strength, while the $h = \pm \half$ 
states come dominantly from the Goldstino.  In the scenario that we are 
studying, the mass of the gravitino is on the scale of keV, while the 
energy with which the gravitino is emitted is on the scale of GeV.  Thus,
it is a very good approximation to ignore the gravitational component and
consider the gravitino purely as a spin $\half$ Goldstino.  From here on,
we will use the symbol $\s G$ to denote the Goldstino.
 
The coupling of one Goldstino to matter is given by the coupling to the
supercurrent  \cite{lowenergy}
\beq
   \delta\L =  -{1\over \sqrt{2} F} \del_\mu \s G c J^\mu
  + {1\over \sqrt{2} F} J^{\mu\dagger} c \del_\mu G^* \ ,
\eeq{LforG}
where $\sqrt{F}$ is the scale of supersymmetry breaking and 
$c = -i \sigma^2$.  The supercurrent  
takes the form
\beqa
   J^\mu & = & \sqrt{2} \sigma^\nu\bar\sigma^\mu D_\nu \phi^* \psi 
   - \sqrt{2}i \left({\del W\over \del \phi}\right)^* \sigma^\mu c \psi^* \CR
         &  & - g\sigma^\mu c \phi^* \lambda^* \phi - i
   \sigma^{\lambda \sigma}F_{\lambda \sigma} \sigma^\mu c \lambda^* \ ,
\eeqa{formofJ}
summed over all chiral supermultiplets $(\phi,\psi)$ and all gauge 
supermultiplets $(A_\mu,\lambda)$. In this equation, $W$ is the superpotential
and $g$ is the gauge coupling.
All of the various terms in this equation actually enter the amplitude for the
three-body stop decay.

It is a formidable task to present the complete dependence of the properties
of the three-body stop decay on the various supersymmetry parameters.
We will present results in this paper for the following four scenarios, 
which illustrate the range of possibilities for the wino-Higgsino mixing 
problem:
\begin{enumerate}
\item a scenario in which the lightest chargino is light and wino-like:
    $m_2 = 200$ GeV, $\mu = 1000$ GeV,
\item a scenario in which the lightest chargino is light and Higgsino-like:
    $m_2 = 1000$ GeV, $\mu = 200$ GeV,
\item a scenario in which the lightest chargino is light and mixed:
    $m_2 = \mu = 260$ GeV,
\item a scenario in which the lightest chargino is heavy:
      $m_2 = \mu = 500$ GeV.
\end{enumerate}
Within each scenario, we will vary other parameters such as $m$,
 $\sin\theta_t$, and $\tan\beta$ in order to gain a more complete picture
of the $\s t_1$ decay.

\section{Characteristics of the stop decay}

\begin{figure}
\begin{center}
\leavevmode
{\epsfxsize=5.00truein \epsfbox{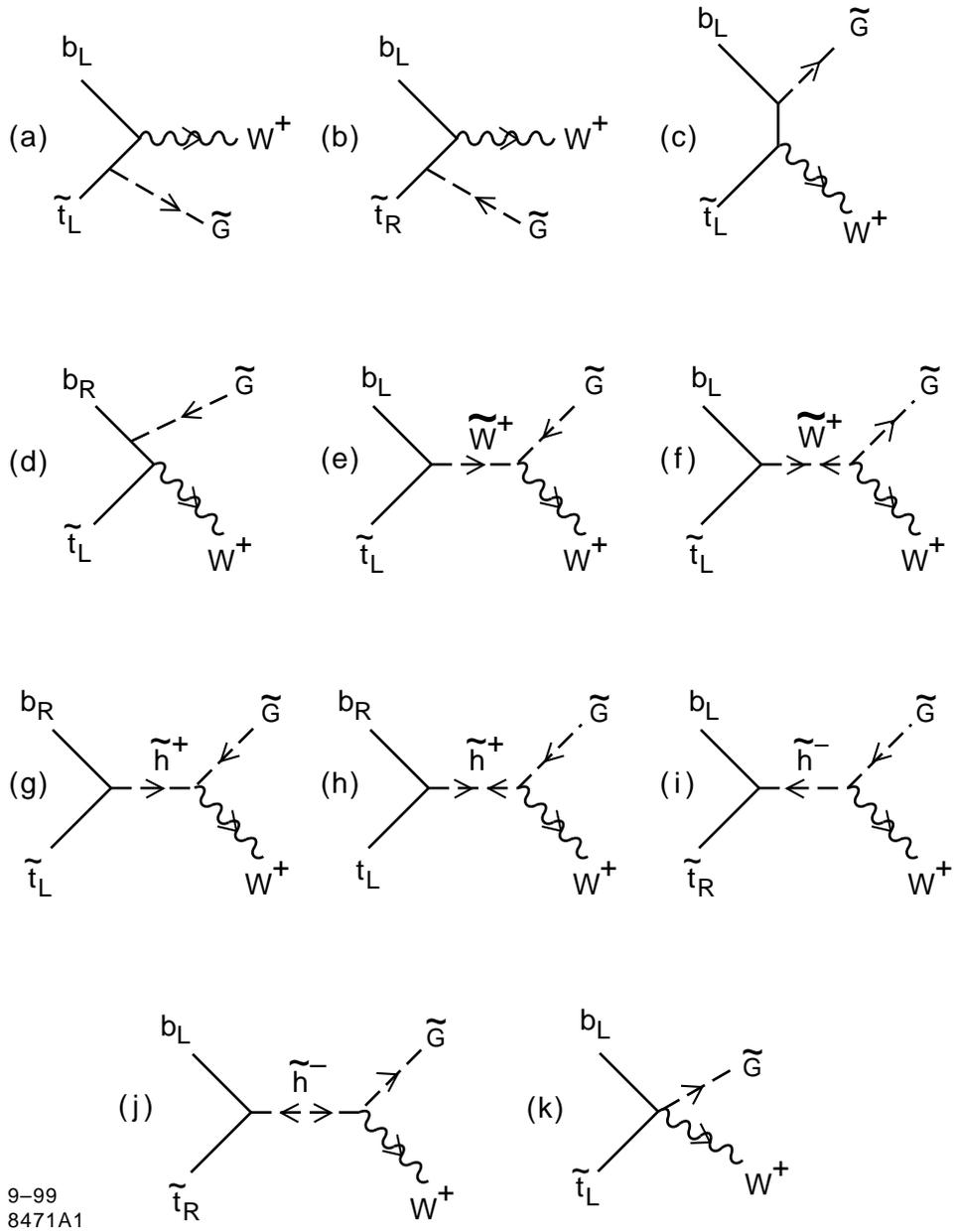}}
\end{center}
\caption{Feynman diagrams for the process ${\s t_1}
   \to  bW{\s G}$. All diagrams are drawn in
 terms of 2-component fermion notation. $\tilde G$ denotes the 
 Goldstino.  The label on the $\s W$/$\s h$ internal lines labels the 
 vertex with which the chargino couples to the top quark.}
\label{fig:Feynman}
\end{figure}

Using the Goldstino interactions from \leqn{LforG} and the gauge and Yukawa
 interactions
of the MSSM, we can construct the Feynman diagrams for $\s t_1 \to
b W^+ \s G$ shown in Figure~\ref{fig:Feynman}.  These
diagrams include processes with intermediate $t$, $\ch{i}$, and $\s b$
particles, plus a contact interaction present in  \leqn{LforG}.

%%%%%%%%%%%%%%%%%%%%%%%%%%%%%%%%%%%%%%%%%%%%%%%%%%%%%%%%%%%%%%%%%%%%%%
%%%%%%%%%%%%%%%%%%%%%%%%%%%%%%%%%%%%%%%%%%%%%%%%%%%%%%%%%%%%%%%%%%%%%%

 It is 
useful to think about building up the complete amplitude for the stop
decay by successively considering a number of limiting cases.
In Figure~\ref{fig:Feynman}, we have drawn the diagrams using a  basis of 
weak interaction eigenstates. 

The first property to be derived from these amplitudes is the stop
decay rate.  It is always an issue when an NLSP decays to the gravitino
whether the decay is prompt on the times scales of particle physics, or
whether the NLSP travels a measureable distance from the production vertex
before decaying. Taking into account the 3-body phase space and
 the fact that the amplitude
is proportional to $1/F$, we might roughly estimate the decay amplitude as
\beq
     \Gamma(\s t_1) \sim {\alpha_w (m - \mw)^7 \over 1028 \pi^2 \mw^2 F^2}\ , 
\eeq{guesstau}
where $\alpha_w = g^2_2/4\pi$ is the weak-interaction coupling constant.
By this estimate, a value of $\sqrt{F}$ smaller than 100 TeV would give a 
prompt decay, with $c\tau < 1$ cm.

%%%%%%%%%%%%%%%%%%%%%%%%%%%%%%%%%%%%%%%%%%%%%%%%%%%%%%%%%%%%%%%%%%%%%%
\begin{figure}
\begin{center}
\leavevmode
{\epsfxsize=4.00truein \epsfbox{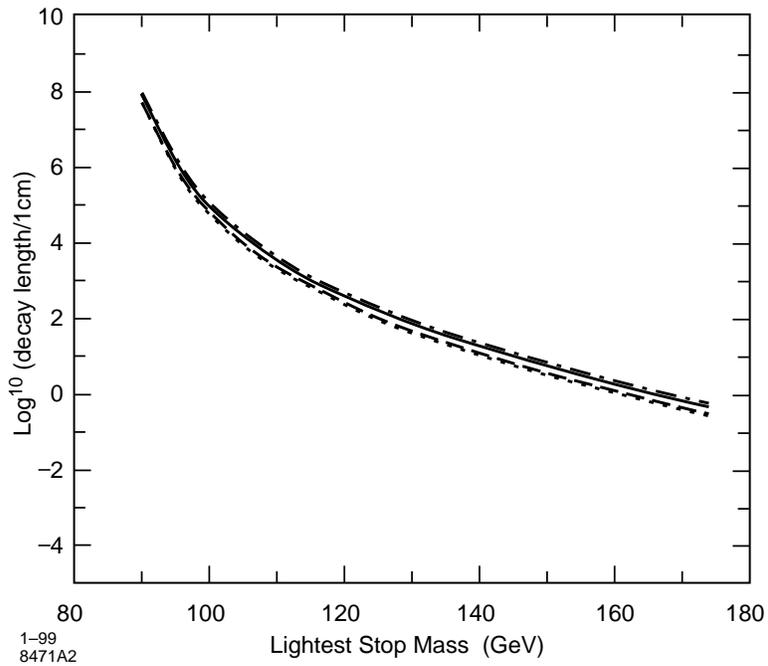}}
\end{center}
\caption{Decay length for the lightest scalar top quark
 $\tilde t_1$ as a function of its  mass 
 in four different scenarios. The
 parameters $\sqrt{F}=30$ TeV, $\tan\beta=1.0$, $m_{\tilde b_L}=300$ GeV, 
and $\sin {\theta_t}=-0.8$ are assumed in all four scenarios.}
\label{fig:DLfig}
\end{figure}
%%%%%%%%%%%%%%%%%%%%%%%%%%%%%%%%%%%%%%%%%%%%%%%%%%%%%%%%%%%%%%%%%%%%%%

In Figure~\ref{fig:DLfig} we show the result of 
a complete calculation of the decay rate in the four scenarios listed at 
the end of Section 2. 
In all four cases, we have chosen the parameter values  $\sqrt F = 30$ TeV,
 $\tan\beta=1.0$, $m_{\tilde b_L}=300$ GeV, and 
$\sin {\theta_t}=-0.8$.  The complete calculation reproduces the steep
dependence on the stop mass $m$ which is present in \leqn{guesstau}.
and shows that the normalization is roughly correct.
 Since $\Gamma$ varies as the fourth
power of $\sqrt F$, one can  arrange for a short decay length by making
$\sqrt F$ sufficiently low.  For $m \sim 160$ GeV, the 
choice  $\sqrt F = 30$
leads to a decay length $c\tau$ of about 1 cm.  We have found that the
decay length is quite insensitive to all of the other relevant parameters.
The variation between scenarios or within a given scenario
 is less than a factor
of 2.  From here on, we will analyze the $\s t_1$ decay  as if it were
prompt.  But it is clear from the figure that, if $\sqrt F$ is as low as 
30 TeV, stop decays will be identifiable by their displaced vertices
in addition to the kinematic signatures discussed in this paper.

The final state of the three-body stop decay is essentially the same as 
ordinary top decay, since the stop produces a $b$ jet, a $W$ boson, and
an unobservable $\s G$.  How, then, can we distinguish the $t \bar t$
and $\s t_1 \s {\bar t}_1$ production processes?  The most straightforward
way to approach this problem is to analyze the observable mass distributions
of $t$ and $\s t_1$ decay products.  If we could completely reconstruct the 
$W$ boson, the invariant mass of the $bW$ system would peak sharply at 
$m_t$ in the case of $t$ decay, and would have a more extended distribution
below the stop mass $m$ in the case of $\s t_1$ decay.  However, in the
observation of top events at the Tevatron, the analysis cannot be so clean.
Events from $t\bar t$ production are typically observed in the final state
in which one $W$ decays hadronically  and the second decays to $\ell \nu$.
  Then
the final state contains an unobserved neutrino.  If there is only this one 
missing particle, the event can be reconstructed.  But the events with 
$\s t_1$ contain two more missing particles, the $\s G$s, which 
potentially confuse the analysis.

\begin{figure}
\begin{center}
\leavevmode
{\epsfxsize=5.00truein \epsfbox{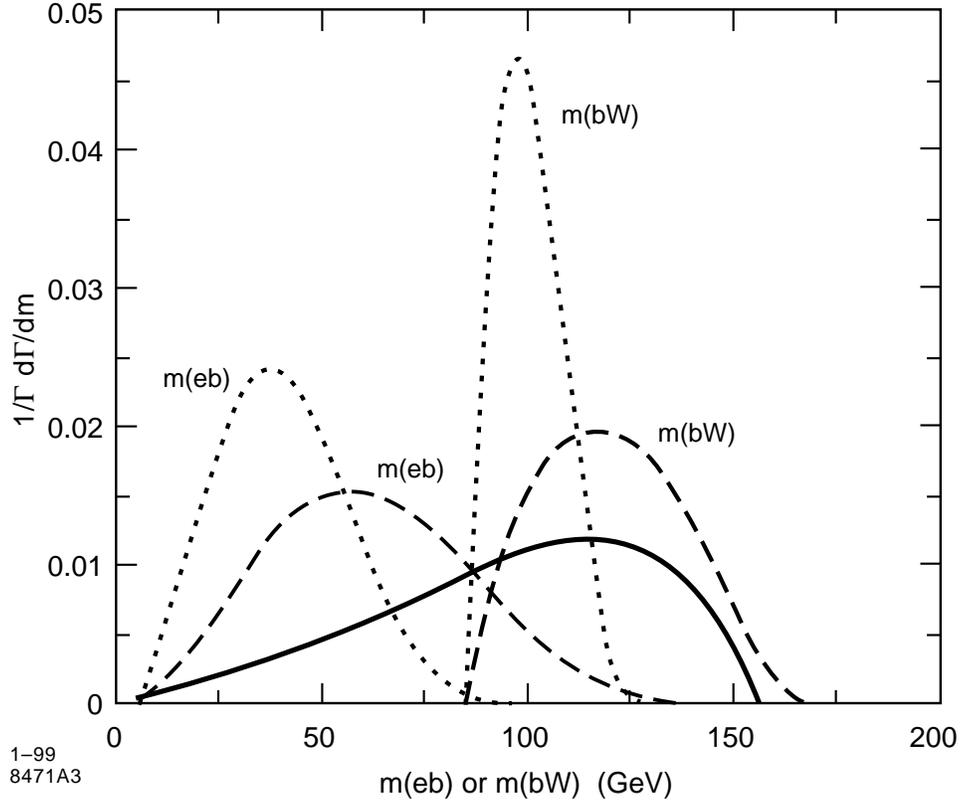}}
\end{center}
\caption{Typical $eb$ and $Wb$ invariant mass distributions for $\s t_1$ decay.
The distributions are shown at two different assumed 
$\s t_1$ masses, $m_{\tilde t_1}=130$ (dotted lines)
 and  170 GeV (dashed lines),
 under the scenario (1). The other parameters are chosen as $\tan\beta=1.0$, 
$m_{\tilde b_L}=300$~GeV, and $\sin {\theta_{t}}=-0.8$. 
The solid line shows, for comparison, the 
$m_{eb}$ spectrum for the standard top quark decay.}
\label{fig:invM}
\end{figure}

Fortunately, it is possible to discriminate $t$ from $\s t_1$ events by 
studying the invariant mass distribution of the directly observable $b$ and
lepton decay products.  For top decays, the distribution in the $b$-lepton
invariant mass $m(eb)$ (quoted, for simplicity, for $m_b = 0$)
takes the form
\beq
    {1\over \Gamma} {d \Gamma \over d m(eb) } =  {12 m(eb)\over
     2 (1-\mw^2/\mt^2)(2 +\mw^2/\mt^2)}(1 - y) \bigl(1 - y +
    {\mt^2\over 2\mw^2}y\bigr)\ , 
\eeq{topbl}
where $y = m^2(eb)/(\mt^2 - \mw^2)$.
As is shown in Figure~\ref{fig:invM},
this distribution extends from $m(eb) = m_b$ to a kinematic endpoint
at $m(eb) = 155$ GeV, and peaks toward  its  high end, at about 
$m(eb) = 120$ GeV.  On the other hand, in $\s t_1$ decay, not only does 
the $m(eb)$ distribution have a lower endpoint value, reflecting the 
value of $m < m_t$, but it also peaks toward the low end of its range.
Figure~\ref{fig:invM} shows two typical distributions of $m(eb)$, 
corresponding to stop masses of 130 and 170 GeV.  The corresponding 
distributions of the $b$-$W$ invariant mass $m(bW)$ are also shown for
comparison. 

%%%%%%%%%%%%%%%%%%%%%%%%%%%%%%%%%%%%%%%%%%%%%%%%%%%%%%%%%%%%%%%%%%%%%%
%%%%%%%%%%%%%%%%%%%%%%%%%%%%%%%%%%%%%%%%%%%%%%%%%%%%%%%%%%%%%%%%%%%%%%

A remarkable feature of  Figure~\ref{fig:invM} is that the $m(eb)$ 
distributions from top and stop decay remain distinctly different even
in the limit in which the stop mass $m$ approaches $m_t$.  Naively, one
might imagine that the stop decay 
diagrams with top quark poles, (a) and (b) in
 Figure~\ref{fig:Feynman}, would dominate in this limit and cause the stop
decay to resemble top decay.  Instead, we find that the top pole diagrams
have no special importance in this limit.  If $E_G$ is the $\s G$ energy,
the top quark pole gives an energy denominator $1/E_G$, but this is 
cancelled by a $\s G$ emission vertex proportional to $(E_G)^{3/2}$.

%%%%%%%%%%%%%%%%%%%%%%%%%%%%%%%%%%%%%%%%%%%%%%%%%%%%%%%%%%%%%%%%%%%%%%
\begin{figure}
\begin{center}
\leavevmode
{\epsfxsize=4.90truein \epsfbox{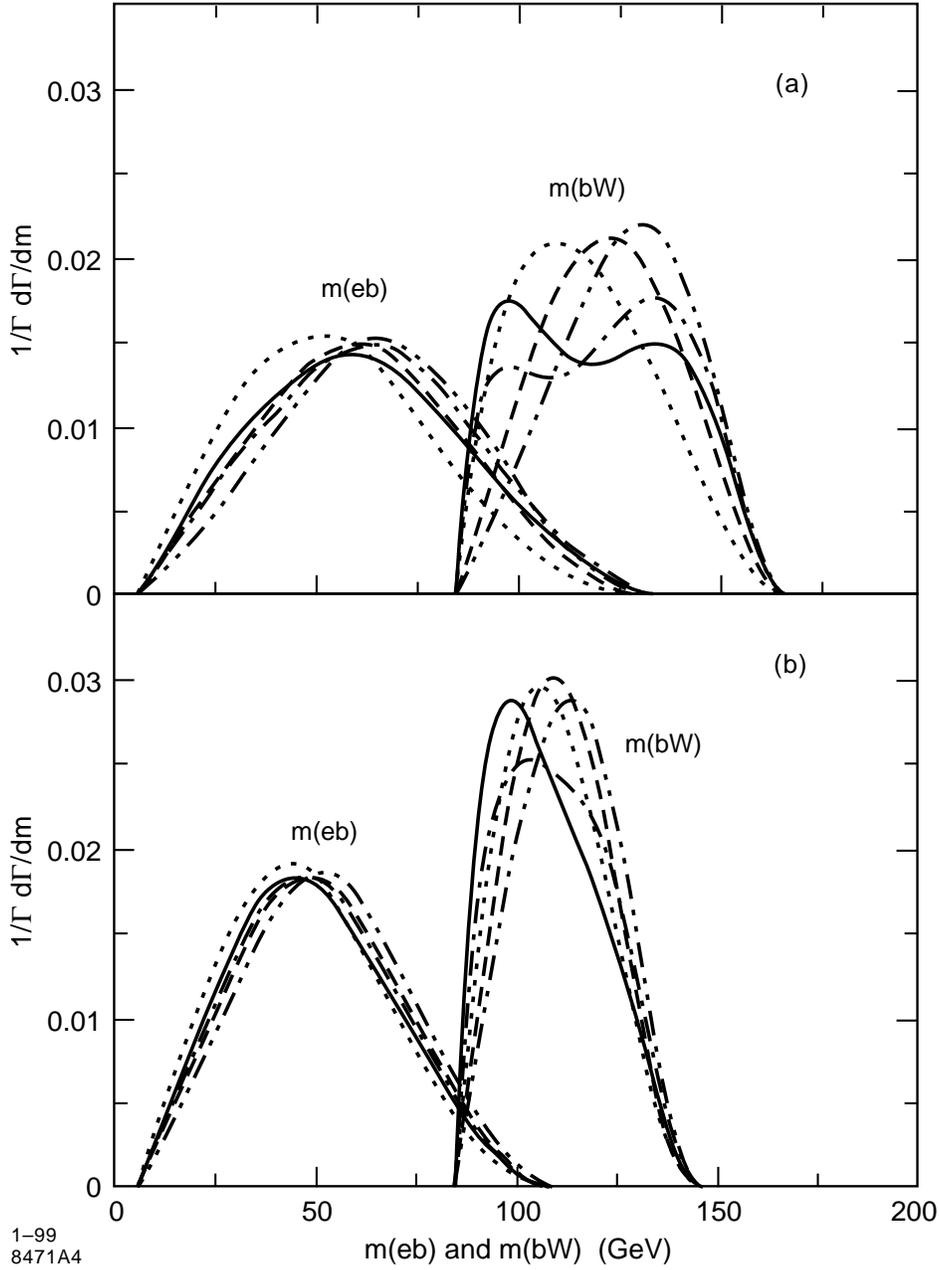}}
\end{center}
\caption{Invariant mass spectra under different scenarios at the two 
mass values
 $m_{\tilde t_1}$: (a)$m_{\tilde t_1}=170$ GeV,  (b) $m_{\tilde t_1}=150$ GeV.
 For each case, the parameters are:
 solid line: scenario (1), $\sin {\theta_{t}}=0.0$, $\tan\beta=1.0$,
 $m_{\tilde b_L}=300$ GeV;  dotted line: scenario (2),
 $\sin {\theta_{t}}=-0.8$,
 $\tan\beta=1.0$, $m_{\tilde b_L}=300$ GeV;  dashed line: scenario (3), 
 $\sin {\theta_{t}}=0.9$, $\tan\beta=50.0$, 
$m_{\tilde b_L}=200$~GeV; dot-dashed line: scenario (3),
 $\sin {\theta_{t}}=0.0$, $\tan\beta=1.0$, 
$m_{\tilde b_L}=300$ GeV;
long-dashed line: scenario (4),
$\sin {\theta_{t}}=0.4$, $\tan\beta=8.0$, $m_{\tilde b_L}=200$ GeV.}
\label{fig:invMvary}
\end{figure}
%%%%%%%%%%%%%%%%%%%%%%%%%%%%%%%%%%%%%%%%%%%%%%%%%%%%%%%%%%%%%%%%%%%%%%

In Figure~\ref{fig:invMvary}, we show the variation of the 
 the distribution of $m(eb)$ and $m(bW)$
 according to the choice of the supersymmetry 
parameters.  The five curves in each group correspond specific parameter
choices in 
the four 
scenarios listed at the end of Section 2, plus an additional choice in 
scenario (3) corresponding to the case of a pure $\s t_L$ ($\theta_t = 0$).
The distributions for a given value of $m$ are remarkably similar. 
Presumably, the shape of these distributions is determined more by 
general kinematic constraints than by the details of the decay amplitudes.
The only exception to this rule that we have found comes in the case
where the $\s t_1$ is dominantly $\s t_L$ and the $\s w$ exchange process
is especially important.

From these results, we believe that the $\s t_1$ production process can be
identified by measuring the distribution of $m(eb)$ in events that pass
the top quark selection criteria.  The mass of the $\s t_1$ can be estimated
from this distribution to about 5 GeV without further knowledge of the 
other supersymmetry parameters.

\section{Longitudinal $W$ polarization}

One of the characteristic predictions of the standard model for top decay
is that the final-state  $W$ bosons  should be highly longitudinally 
polarized.  Define the degree of longitudinal polarization by 
\beq
  r =  {\Gamma (W_0) \over \Gamma(\mbox{all})} \ .
\eeq{rdef}
Then the leading-order prediction for this polarization  in top decay is
\beq 
r_t = {1 \over 1 + 2 \mw^2/ m_t^2} \approx 0.71  \ .
\eeq{topratio}
We have seen already that the configuration  of the final $b W^+$ system
in stop decay is quite different from that in top decay.  Thus, it would
seem likely that the longitudinal $W$ polarization would also deviate from 
the characteristic values for top.  We will show that the value of $r$ in
stop decay typically differs significantly from \leqn{topratio}, in a manner
that gives information about the underlying supersymmetry model.

The measurement of the polarization $r$ at the Tevatron has been studied
using the technique of  reconstructing the $W$ decay angle 
in single-lepton events from the lepton and neutrino 
four-vectors \cite{CDF,TeV33}.
An accuracy of $\pm 0.03$ should be achieved in the upcoming Run II.
This technique, however, cannot be used for stop events, since the missing
momentum includes the $\tilde G$'s as well as the neutrino.
However, one can also measure the longitudinal $W$ polarization from 
the $W$ decay angle determined by
using the four-vectors of the two jets assigned to the hadronic $W$ in the
event reconstruction. It is not necessary to distinguish the quark from 
the antiquark to determine the degree of longitudinal polarization.  

\begin{figure}
\begin{center}
\leavevmode
{\epsfxsize=4.00truein \epsfbox{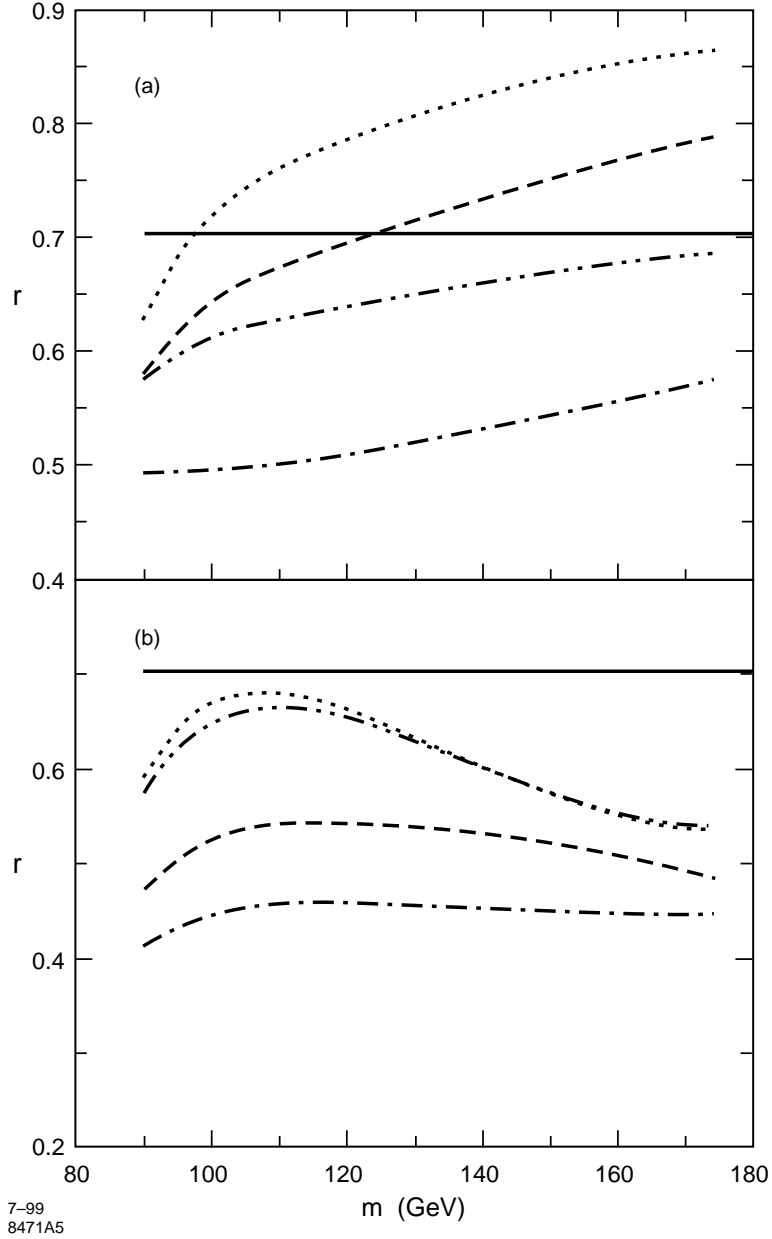}}
\end{center}
\caption{Longitudinal W production ratio for the stop $\tilde t_1$
 decay as a function of $m_{\s t_1}$ under different scenarios.
  The dot-dash, dotted, dashed, and dot-dot-dash lines refer, respectively,
 to the chargino scenarios (1), (2), (3), (4) given at the end of 
 Section 2.  The two figures show
 (a)  $\tilde t_R$-like cases with $sin{\theta_{t}}=-0.8$, 
$tan\beta=1.0$, and $m_{\tilde b_L}=300$ GeV. (b) pure
 $\tilde t_L$-like cases with $sin {\theta_{t}}=0.0$,
 $tan\beta=1.0$, and $m_{\tilde b_L}=300$ GeV. }
\label{fig:Whel}
\end{figure}

\begin{figure}
\begin{center}
\leavevmode
{\epsfxsize=4.00truein \epsfbox{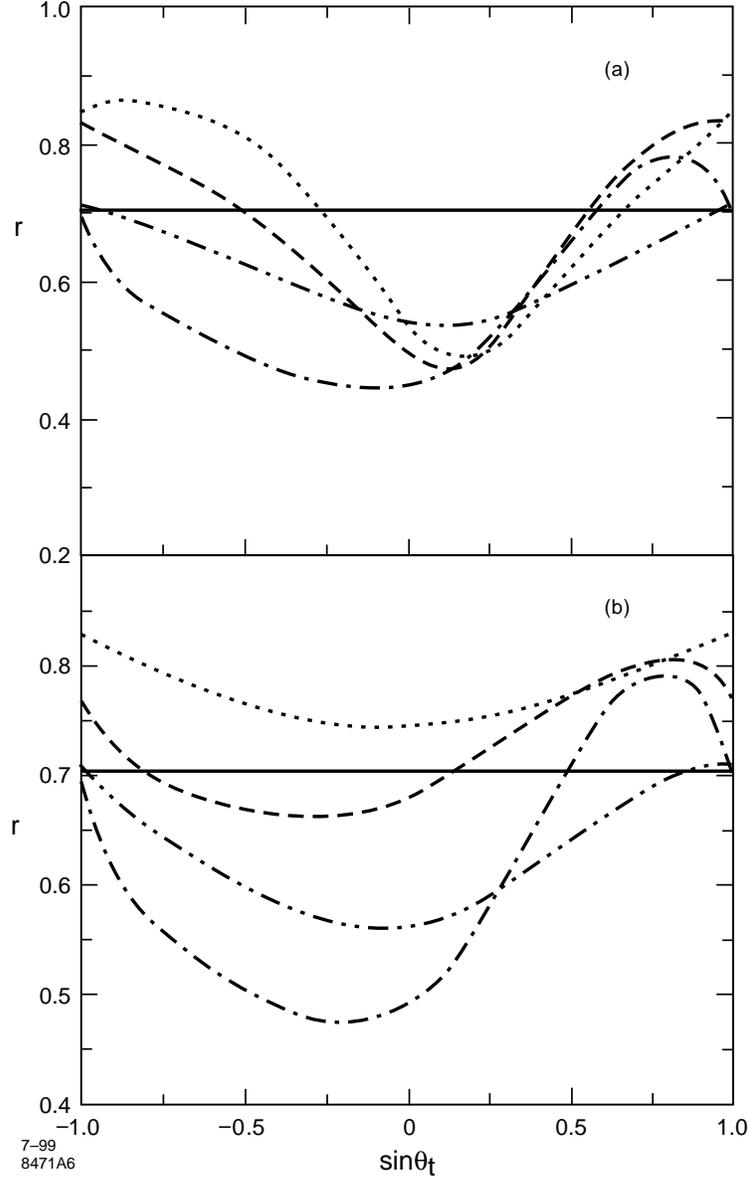}}
\end{center}
\caption{Longitudinal W production ratio for the stop $\tilde t_1$ decay
 as a function of $sin {\theta_{t}}$ under different scenarios.
  The dot-dash, dotted, dashed, and dot-dot-dash lines refer, respectively,
 to the chargino scenarios (1), (2), (3), (4) given at the end of 
 Section 2.  The two figures show the dependence for 
 (a) $m_{\tilde t_1}=170$~GeV, $tan\beta=1.0$, and $m_{\tilde b_L}=300$ GeV.
 (b) $m_{\tilde t_1}=170$~GeV, $tan\beta=50.0$, and $m_{\tilde b_L}=300$ GeV.}
\label{fig:WhelSin}
\end{figure}

What value of $r$ should be found for light stop pair production?  In 
Figures~\ref{fig:Whel} and~\ref{fig:WhelSin}, we plot the value of $r$ in 
the four scenarios listed at the end of Section 2, for representative 
values of the parameters, as a function of the stop mass.  We see that the 
value of $r$ is typically lower than the top quark value \leqn{topratio},
that it has a  slow dependence on the value of the stop mass $m$, and that
it can depend significantly on the stop mixing angle $\theta_t$.

%%%%%%%%%%%%%%%%%%%%%%%%%%%%%%%%%%%%%%%%%%%%%%%%%%%%%%%%%%%%%%%%%%%%%%
%%%%%%%%%%%%%%%%%%%%%%%%%%%%%%%%%%%%%%%%%%%%%%%%%%%%%%%%%%%%%%%%%%%%%%
%%%%%%%%%%%%%%%%%%%%%%%%%%%%%%%%%%%%%%%%%%%%%%%%%%%%%%%%%%%%%%%%%%%%%%
%%%%%%%%%%%%%%%%%%%%%%%%%%%%%%%%%%%%%%%%%%%%%%%%%%%%%%%%%%%%%%%%%%%%%%

%%%%%%%%%%%%%%%%%%%%%%%%%%%%%%%%%%%%%%%%%%%%%%%%%%%%%%%%%%%%%%%%%%%%%%
\begin{figure}
\begin{center}
\leavevmode
{\epsfxsize=4.00truein \epsfbox{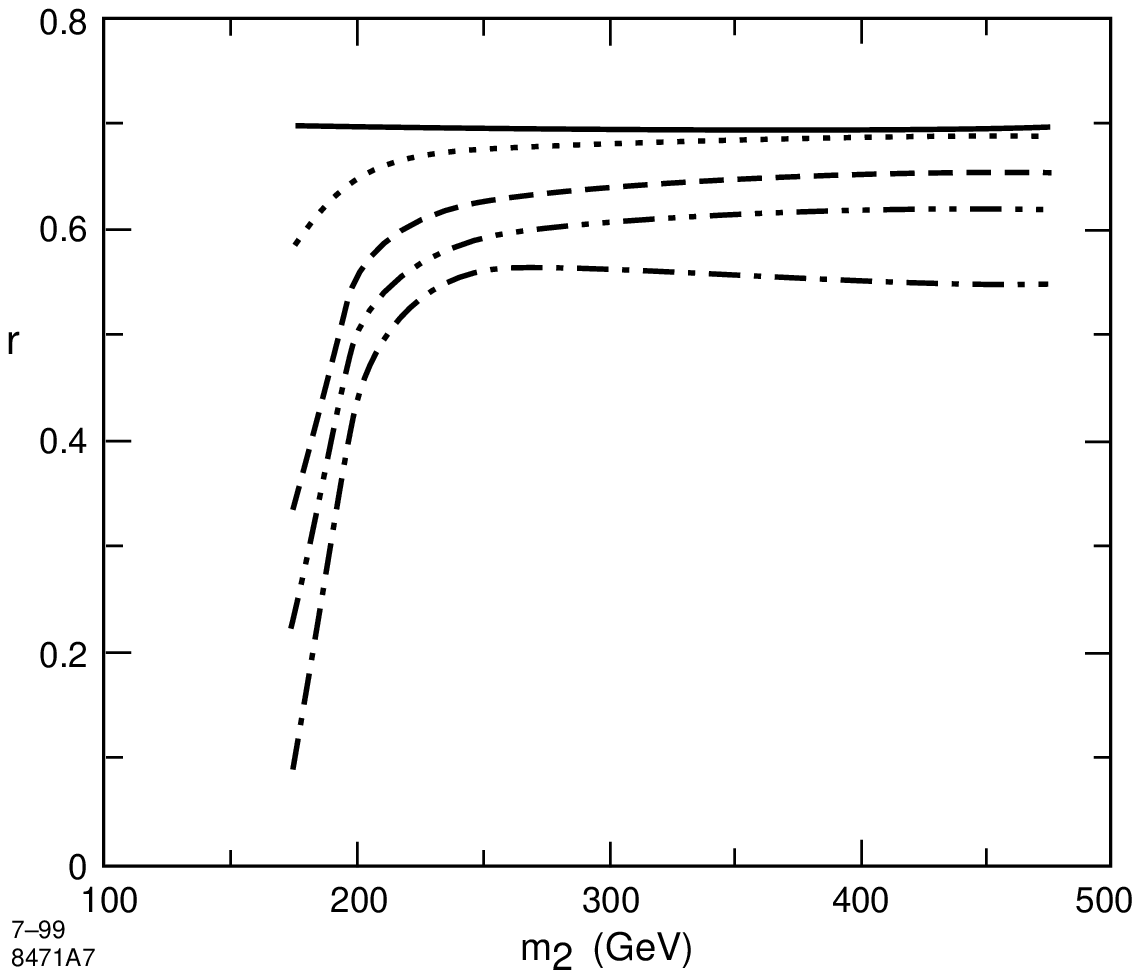}}
\end{center}
\caption{Longitudinal W production ratio for the stop $\tilde t_1$
 decay as a function of the soft breaking mass of SU(2) gaugino.  The
four curves correspond to $\sin\theta_t =$ -0.98, -0.8, -0.6, 0.0.
 The other parameters are chosen to be 
 $m_{\tilde t_1}=170$ GeV, $tan\beta=1.0$, $\mu=1000$ GeV, and
 $m_{\tilde b_L}=300$ GeV.}
\label{fig:WhelM2}
\end{figure}
%%%%%%%%%%%%%%%%%%%%%%%%%%%%%%%%%%%%%%%%%%%%%%%%%%%%%%%%%%%%%%%%%%%%%%

The variation of $r$ arises from the competition between the diagrams in 
Figure~\ref{fig:Feynman} in which the Goldstino is radiated from the $t$ and
$b$ legs and those in which the Goldstino is radiated from the $W$. To 
understand this, it is 
useful to think about the limiting cases  in which each intermediate 
propagator goes on shell.  In the case in which the top quark goes on 
shell in diagrams a,b of Figure~\ref{fig:Feynman}, the $W$ polarization has
the same vlaue \leqn{topratio} as that for top decay.  In the case in which 
the $\s b$ goes on shell, we have the process $\s t \to \s b W^+$, for
which also $r = 1/(1+ 2 \mw^2/\mt^2)$.  However, the third case in which
the $\s \chi^+$ goes on shell can give a very different result.  In the limit
in which the $\s \chi^+$ is pure gaugino, we have the subprocess 
$\s w^+ \to \s G W^+$, which leads to purely transversely polarized $W$ 
bosons.  More generally, for the process  $\ch{1} \to \s G W^+$ on shell,
we have
\beq
     r =  { |V_{+12}|^2 + |V_{-12}|^2 \over 2 (  |V_{+11}|^2 + |V_{-11}|^2 )
                   +  |V_{+12}|^2 + |V_{-12}|^2 } \ ,
\eeq{charginor}
where $V_+$, $V_-$ are the matrices defined in \leqn{chdiag}.
These individual components vary in importance as the masses on the 
intermediate lines are varied.   The role of the chargino diagrams
in producing a low value of $r$
is shown clearly in 
Figure~\ref{fig:WhelM2}.  Here we plot the value of $r$ as a function of
the supersymmetry-breaking $SU(2)$ gaugino mass $m_2$ and observe that 
$r$ moves to a higher asymptotic value as the gaugino is decoupled.

Beyond this observation, though, the dependence of $r$ on the underlying
parameters is not simple.  As we have seen in the previous section,
 it is never true that one particular subprocess comes almost onto
mass shell and dominates the stop decay.   This feature of the stop decay,
which was an advantage in the previous section, here provides a barrier to 
finding quantitative relation between a measured value of $r$ and the
underlying parameter set.   On the other hand, it is interesting that almost
every scenario predicts a value of  $r$ substantially different
from the Standard Model value for top decay.

\section{Conclusions}

In this paper, we have discussed the phenomenology of light stop decay through
the process $\s t \to W^+ b \s G$.  We have shown that this process can be
distinguished from $t$ decay through the characteristic shape of the 
$b\ell$ mass distribution.  We have shown also that the fraction of 
longitudinal polarization of the 
$W^+$
in $\s t$ decays can vary significantly from the prediction \leqn{topratio}
for $t$.  Since these two observables are available at the Tevatron collider,
it should be possible there to exclude or confirm this unusual scenario
for the realization of supersymmetry.

\bigskip\bigskip

  \Acknowledgements

We are grateful to Scott Thomas for suggesting this problem, to Regina
Demina, for encouragement and discussions on stop experimentation, and to
JoAnne Hewett, for helpful advice.  This work was supported by the
Department of Energy under contract  DE--AC03--76SF00515.

%%%%%%%%%%%%%%%%%%%%%%%%%%%%%%%%%%%%%%%%%%%%%%%%%%%%%%%%%%%%%%%%%%%%%%%

%%%%%%%%%%%%%%%%%%%%%%%%%%%%%%%%%%%%%%%%%%%%%%%%%%%%%%%%%%%%%%%%%%%%%%%%%
%\newpage
%\listoffigures

\end{document}